\documentclass[a4paper,11pt]{article}
\usepackage{amsmath,amssymb,amsfonts,amsthm,latexsym} 

\usepackage{pos}
\usepackage[normalem]{ulem}
\usepackage[utf8]{inputenc}
\usepackage[T1]{fontenc}
\usepackage{cancel}
\usepackage{graphicx}
\usepackage{latexsym}
\usepackage{bbm}
\usepackage{bm}
\usepackage{epsfig}
\usepackage{epstopdf}
\usepackage{subcaption}
\usepackage{color}
\usepackage{comment}
\usepackage{slashed}
\usepackage{placeins}
\usepackage{cleveref}
\usepackage{setspace}
\usepackage{url}
\usepackage{tabularx}
\usepackage{enumitem}
\usepackage{hyperref}
\usepackage{diagbox}
\usepackage{dcolumn}
\usepackage{xfrac}
\usepackage{pgf}

\newcommand{\tr}{\mathop{\mathrm{tr}}}
\newcommand{\dd}[1]{\mathop{\mathrm{d}#1}}
\newcommand{\Nf}{N_{\rm f}}
\newcommand{\Nc}{N_{\rm c}}
\newcommand{\Ns}{N_{\rm s}}
\newcommand{\Nt}{N_{\rm t}}
\newcommand{\CA}{C_{\rm A}}

\newcommand{\CF}{C_{\rm F}}

\newcommand{\e}[1]{\text{e}^{#1}}

\newcommand{\as}{\alpha_{\mathrm{s}}}
\newcommand{\als}{\alpha_{\mathrm{s}}}

\newcommand{\mc}{m_{\text{c}}}

\newcommand{\MSb}{\overline{\textrm{MS}}}
\newcommand{\QCD}{{\MSb}}

\newcommand{\be}{\begin{equation}} 
\newcommand{\ee}{\end{equation}}
\def\ba#1\ea{\begin{align}#1\end{align}}
\newcommand{\bea}{\begin{eqnarray}} 
\newcommand{\eea}{\end{eqnarray}}
\def\lsim{\mathrel{\raise.3ex\hbox{$<$\kern-.75em\lower1ex\hbox{$\sim$}}}}
\def\gsim{\mathrel{\raise.3ex\hbox{$>$\kern-.75em\lower1ex\hbox{$\sim$}}}}

\renewcommand{\sfrac}[2]{\ensuremath{#1/#2}}

\renewcommand{\sfrac}[2]{\ensuremath{#1/#2}}
\newcommand{\Nstates}{\ensuremath{N_{\text{st}}}}

\colorlet{preliminary}{White!92!BrickRed}

\renewenvironment{thebibliography}[1]{
  \begin{oldthebibliography}{#1}
    \setlength{\itemsep}{0em}
    \setlength{\parskip}{0em}
}
{
  \end{oldthebibliography}
}
\let\oldcite\cite
\renewcommand{\cite}[1]{\mbox{\oldcite{#1}}}

\title{Strong coupling in (2+1+1)-flavor QCD}
\author*[a,b]{Viljami~Leino}
\author[c]{Alexei~Bazavov}
\author[d,e,f]{Nora~Brambilla}
\author[g,f]{Andreas~S.~Kronfeld}
\author[d,e]{Julian~Mayer-Steudte}
\author[i]{Peter~Petreczky}
\author[d,j]{Sebastian~Steinbeißer}
\author[d]{Antonio~Vairo}
\author[k,l]{Johannes~H.~Weber}

\affiliation[a]{Helmholtz Institute Mainz, Staudingerweg 18, 55128 Mainz, Germany}
\affiliation[b]{Institut für Kernphysik, Johannes Gutenberg-Universität Mainz,\\
             Johann-Joachim-Becher-Weg 48, 55128 Mainz, Germany}
\affiliation[c]{Department of Computational Mathematics, Science and Engineering, and Department of \\
Physics and Astronomy, Michigan State University, East Lansing, Michigan 48824, USA} 
\affiliation[d]{Physics Department, TUM School of Natural Sciences, 
Technical University of Munich, 
\\ 
James-Franck-Straße~1, 85748 Garching b.\ München, Germany}
\affiliation[e]{Munich Data Science Institute, Technical University of Munich, \\ 
Walther-von-Dyck-Straße~10, 85748 Garching b.\ München, Germany}
\affiliation[f]{Institute for Advanced Study, Technical University of Munich, \\ 
Lichtenbergstraße~2a, 85748 Garching b.\ München, Germany}
\affiliation[g]{Particle Theory Department, Theory Division, Fermi National Accelerator Laboratory, \\
    Batavia, Illinois 60510-5011, USA}
\affiliation[i]{Physics Department, Brookhaven National Laboratory, Upton, New York 11973-5000, USA}
\affiliation[j]{Leibniz-Rechenzentrum der Bayerischen Akademie der Wissenschaften, \\ 
Boltzmannstraße~1, 85748 Garching b.\ München, Germany}
\affiliation[k]{Institut f\"ur Physik \& IRIS Adlershof, Humboldt-Universit\"at zu Berlin, 
Zum Gro\ss en Windkanal 2, D-12489 Berlin, Germany}
\affiliation[l]{Institut f\"ur Kernphysik, Technische Universit\"at Darmstadt, 
Schlossgartenstra\ss e 2, D-64289 Darmstadt, Germany}

\emailAdd{viljami.leino@uni-mainz.de}

\abstract{
\vspace*{-1mm}
\textbf{\textsf{TUMQCD Collaboration}}\\[0.25em]
The strong coupling $\as$ can be obtained from the static energy as shown in previous lattices studies. 
For short distances, the static energy can be calculated both on the lattice with the use of Wilson line correlators, and with the perturbation theory up to three loop accuracy with leading ultrasoft log resummation. 
Comparing the perturbative expression and lattice data allows for precise determination of $\as(m_Z)$. 
We will present preliminary results for the determination of $\as {(M_Z)}$ in (2+1+1)-flavor QCD using the configurations made available 
by the MILC-collaboration with smallest lattice spacing reaching 0.0321~fm.}

\FullConference{The 41st International Symposium on Lattice Field Theory (LATTICE2024)\\
 28 July - 3 August 2024\\
Liverpool, UK\\}

\begin{document}
\begin{flushright}
\scriptsize \texttt{FERMILAB-CONF-25-0051-T\\ \vspace{-0.4em} MITP-25-008\\ \vspace{-0.9em} TUM-EFT 195/25}  \normalsize
\end{flushright}
\maketitle

\section{Introduction}
The strong coupling $\as$ is a fundamental parameter of QCD and the {Standard Model} of particle physics.
The running of the strong coupling {in the $\overline{\text{MS}}$ scheme} is a function of the renormalization scale $\mu$ and the intrinsic scale of QCD,
$\Lambda_{\QCD}$.
When $\Lambda_{\QCD}$ is known, one can perturbatively determine $\as$ at any scale $\mu\gg\Lambda_{\QCD}$.
In these proceedings we focus on determining this intrinsic scale.

Observables that can be calculated with high precision in both perturbative- and lattice-QCD are ideal candidates 
to determine $\as$ in the regions where both approaches agree. 
One such observable is the energy between a static quark and a static antiquark separated by distance $r$ known as the static energy $E_0(r)$. 
$E_0(r)$ is a fundamental observable of QCD that played an important role~\cite{Bali:2000gf} in establishing confinement in QCD 
and understanding asymptotic freedom. 
On the lattice, $E_0(r)$ is defined as the ground state of a static Wilson loop. 
At short distances $r\Lambda_{\QCD}\ll 1$, it holds that $\as(1/r) \ll 1$ and the static energy is well defined by a weak coupling perturbative expansion.
This expansion is known up to $\mathrm{N}^3\mathrm{LL}$ level~\cite{Brambilla:1999qa,Pineda:2000gza,Brambilla:2006wp,Brambilla:2009bi,Anzai:2009tm,Smirnov:2009fh}. 
By comparing the perturbative expansion of the static energy to the static energy on the lattice at short distances, we can extract $\Lambda_{\QCD}$. 

So far, $\Lambda_{\MSb}$ has been determined from the static energy in $\Nf=0$ the pure gauge SU(3) Yang-Mills 
theory~\cite{Brambilla:2010pp,Husung:2017qjz,Brambilla:2023fsi} and with either 
$\Nf=2$ dynamical quark flavors~\cite{Jansen:2011vv,Karbstein:2014bsa,Karbstein:2018mzo} or with 
$\Nf=2+1$ dynamical flavors~\cite{Bazavov:2012ka,Bazavov:2014soa,Takaura:2018vcy,Bazavov:2019qoo,Ayala:2020odx}.
With $\Nf=2+1+1$ dynamical flavors, $\Lambda_{\QCD}$ {via the static energy} is yet to be determined, however,
$\Lambda_{\QCD}$ has been determined on the lattice in 2+1+1-flavor QCD with other methods~\cite{Blossier:2013ioa,Chakraborty:2014aca}.
In these proceedings, we report on the progress of the newest TUMQCD lattice extraction of $\Lambda_{\QCD}$ from the lattice with 2+1+1 dynamical {flavors via} the static energy.
For a complete review of the status of $\as$ determined from the lattice QCD, we refer the reader to the recent FLAG review~\cite{Aoki:2024oxs} and for a wider review with also the experimental status to Ref.~\cite{dEnterria:2022hzv}.

\section{Static energy}
\subsection{Perturbation theory}
The static energy $E_0(r)$ has the perturbative expansion
\begin{equation}
    \label{eq:statenergyI}
    E_{0}(r) =\Lambda - \frac{\CF \als}{r}\! \left( 1\! +\! \# \als\! +\! \# \als^{2}\! +\! \# \als^{3} \ln\als\! +\! \# \als^{3}\! +\!
        \# \als^{4} \ln^{2}\als\! +\! \# \als^{4} \ln\als\! + \dots \right),
\end{equation}
where $\Lambda$ is a constant of mass dimension one and the coefficients $\#$ can be found in the review~\cite{Tormo:2013tha}.
The running coupling depends on a soft scale $\mu$ that is commonly set to be $\mu=c/r$ with $c$ varied around 1 to gauge perturbative error.
Starting at order $\as^4$ ultrasoft logarithms are introduced with ultrasoft scale $\nu=\CA\as(\mu)/2r$. 
These logarithms can be resummed, however, in these proceedings we show results only without resummation and leave the study of ultrasoft scale dependence to the final publication.
The expansion given by Eq.~\eqref{eq:statenergyI} is known to the order given for massless sea quarks.
To include the effects from the finite mass of the charm quark, a correction $\delta V_{m}^{(\Nf)}(r)$ has to be added to the static energy.
This correction is known in perturbation theory up to order of $\als^{3}$ (see Ref.~\cite{Recksiegel:2001xq} for summary of equations).
With {(2+1+1)}-flavor QCD, the relevant massive quark is the charm quark that has a mass $\mc^{\MSb}(\mc^{\MSb}) \approx 1.28~\text{GeV} \simeq \sfrac{1}{0.15}~\text{fm}^{-1}$~\cite{Aoki:2024oxs}.

\begin{figure}
    \centering
    \includegraphics[width=0.49\textwidth]{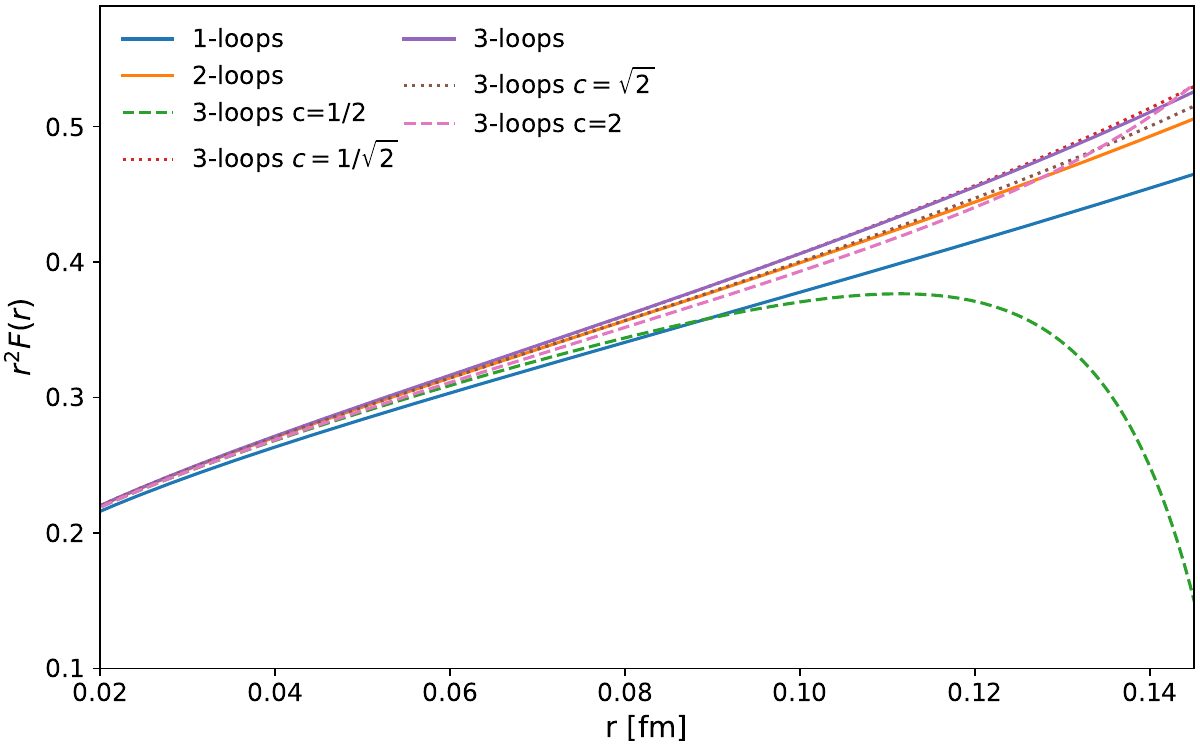}
    \includegraphics[width=0.49\textwidth]{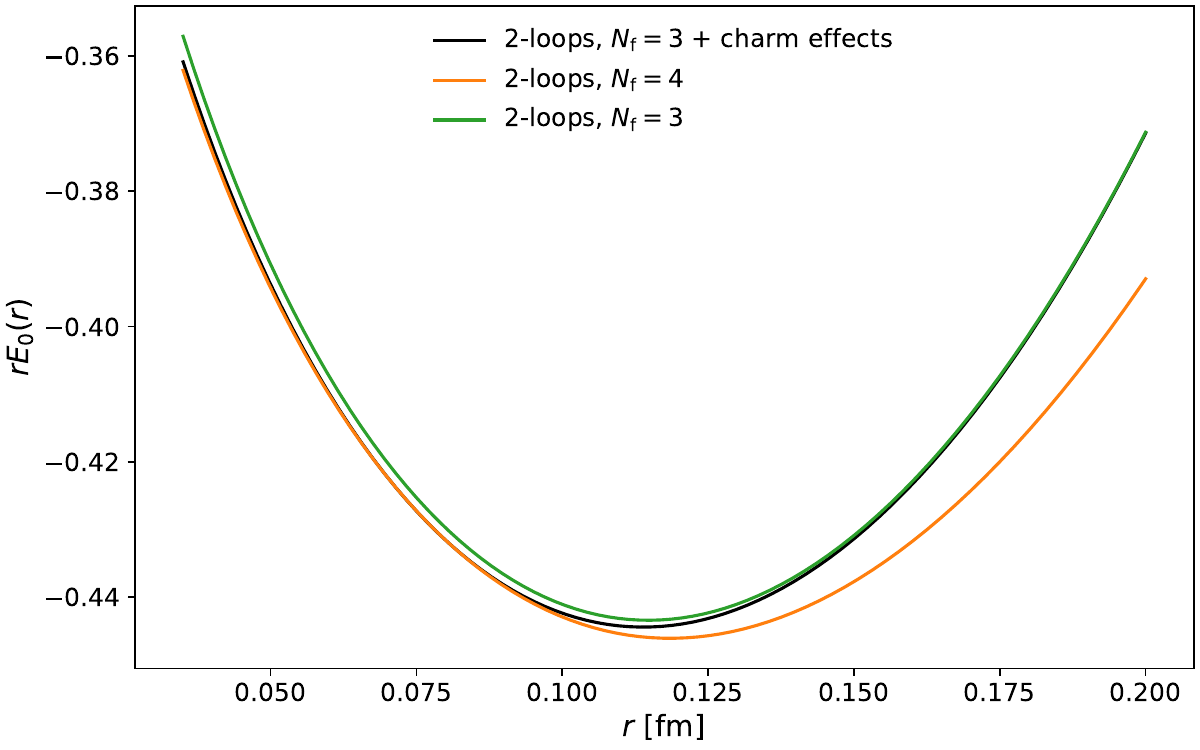}
    \caption{
             Left: The static force $r^2 F(r)$ at different orders of perturbation theory and different scalings of $\mu=c/r$. 
             Right: The static energy at different combinations of light and massive quarks. 
             The free constant shift $\Lambda$ is optimized to minimize the covered range in y-axis to make the differences between curves more visible.
             }
    \label{fig:pert}
\end{figure}
The constant $\Lambda$ in Eq.~(\ref{eq:statenergyI}) is scheme dependent quantity. 
On the lattice it relates to a linear divergence in inverse lattice spacing $a^{-1}$ 
and in dimensional regularization it manifests as a renormalon of mass dimension one.
Since the constant by itself does not depend on $\as$, and all $\as$ dependence is {contained} in the $r$-dependent part of $E_0(r)$,
we can obtain $\as$ from the lattice data by simply matching the lattice results and perturbative curves at some reference distance $r^\ast$.
However, in order to get as stable as possible perturbative behavior, the issue with renormalon needs to be solved. 
We explore two strategies to deal with the renormalon contribution:
\begin{enumerate}
\item The static force $F(r)$, defined as a derivative with respect to $r$ of the static energy, no longer contains the constant $\Lambda$. 
$F(r)$ can be computed directly on the lattice~\cite{Brambilla:2021wqs,Brambilla:2023fsi} and compared to perturbative expansion of the static force to extract $\as$.
However, since we have computed the static energy on the lattice, we take a different approach and integrate the perturbative static force 
as described in~\cite{Bazavov:2014soa,Bazavov:2019qoo} to get more stable definition of $E_0(r)$. 
The constant $\Lambda$ reappears, now as an integration constant, and can be matched to lattice data at distance $r^\ast$. 
The total static {energy} with the finite mass correction then takes the form:
\begin{equation}
    \label{eq:full_statenergy}
    E^{(\Nf)}_{0,m}(r) = \int\limits_{r^{\ast}}^{r} \dd{r^\prime} \; F^{(\Nf)}(r^\prime) + \delta V_{m}^{(\Nf)}(r) + \text{const}\,,
\end{equation}
where $\Nf=3$ is the number of massless quarks. 
In the limit $m \gg \sfrac{1}{r}$, Eq.~\eqref{eq:full_statenergy} reduces to the case of $\Nf$ massless quarks $E^{(\Nf)}_{0}(r)$,
while in the limit $m \ll \sfrac{1}{r}$ it reduces to case of $\Nf+1$ massless quarks $E^{(\Nf+1)}_{0}(r)$.
This is a consequence of the decoupling of charm quark in the static potential.
We demonstrate the entire procedure in figure~\ref{fig:pert}. 
On the left side, we show the static force multiplied by squared distance $r^2F(r)$ 
at different orders of perturbation theory and at different renormalization scales $\mu=c/r$.
On the right side, we show the effect of the charm quark after the force has been integrated back to static energy. 

\item Alternatively, we can inspect the minimal renormalon subtraction (MRS) prescription~\cite{Brambilla:2017hcq,Komijani:2017vep,Kronfeld:2023jab}. 
In this method the leading factorial growth of the expansion coefficients is summed to all orders, which stabilizes the behavior of $\Lambda$ 
and reduces the perturbative error of the fits. 
This approach avoids the need to numerically integrate the static force. 
To deal with the {charm sea}, we add the correction $\delta V_{m}^{(\Nf)}(r)$ at fixed order to the MRS description of $E_0(r)$. 
\end{enumerate}

\subsection{Lattice}
We compute the static energy from the (2+1+1)-flavor lattice ensembles generated by the MILC Collaboration~\cite{MILC:2010pul,MILC:2012znn,Bazavov:2017lyh}. 
For gluons the one-loop Symanzik-improved action with tadpole improvement has been used. 
The sea quarks, namely two isospin-symmetric light quarks, and physical strange and charm quarks, are simulated with the HISQ-action~\cite{Follana:2006rc}.
While the total dataset spans lattice spacings from  0.032~fm to 0.15~fm with light quark mass $m_l/m_s$ being either $1/10$, $1/5$ or physical,
in these proceedings we mainly focus on the finest lattice with lattice spacing $a=0.03216$~fm and $m_l/m_s=1/5$. 
Analysis for many of the coarser ensembles and the continuum extrapolation is left for the final publication.

The gauge configurations have been fixed to Coulomb gauge, which allows for easy access to off-axis distances.
Instead of the Wilson loops, in Coulomb gauge, $E_0(r)$ can be obtained from the time dependence of the Wilson-line correlation function
$C\left(\bm{r},\tau,a\right)$ at separation $\sfrac{\bm{r}}{a}$:
\begin{align}
    C\left(\bm{r},\tau,a\right) &= \left\langle \frac{1}{\Ns^{3}} \sum\limits_{\bm{x}} \sum\limits_{\bm{y}=R(\bm{r})}
        \frac{1}{\Nc\,N_{\bm{r}}} \tr\left[W^{\dagger}\left(\bm{x}+\bm{y},\tau,a\right)
            W\left(\bm{x},\tau,a\right)\right] \right\rangle
    \nonumber \\
        &= \sum_{n=0}^{\infty} C_{n}\left(\bm{r},a\right) \left(\e{-\tau E_{n}\left(\bm{r},a\right)} +
            \e{-(a\Nt-\tau)E_{n}\left(\bm{r},a\right)} \right)
    \nonumber \\
     &= \e{-\tau E_{0}\left(\bm{r},a\right)} \left( C_{0}\left(\bm{r},a\right) +
        \sum\limits_{n=1}^{\Nstates-1} C_{n}\left(\bm{r},a\right)
            \prod\limits_{m=1}^{n} \e{-\tau \Delta_{m}\left(\bm{r},a\right)}
        \right) + \ldots,
    \label{eq:forward_tower}
\end{align}
where we have reparameterized the correlation in terms of energy differences $a\Delta_{n}(\bm{r},a)=aE_{n}(\bm{r},a)-aE_{(n-1)}(\bm{r},a)>0$.
We choose $\Nstates=1$, $2$, or~$3$ to fit our data to this form using Bayesian priors and extract $E_0(r)$.
The reader is referred to our previous publication for further technical details~\cite{Brambilla:2022het} of the procedure. 
In Ref.~\cite{Brambilla:2022het}, we determined the potential scales $r_i$, especially $r_1\simeq0.3~\text{fm}$~\cite{Bernard:2000gd} which is defined from the static force as $r^2_1F(r_1) = 1$. 
We will use the $r_1$-scale as our main scale for the $\Lambda_\mathrm{\MSb}$ extraction. 
The conversion of the results from $r_1$-units to physical units is discussed further in section~\ref{sec:scale}.

\begin{figure}
    \centering
    \includegraphics[width=0.49\textwidth]{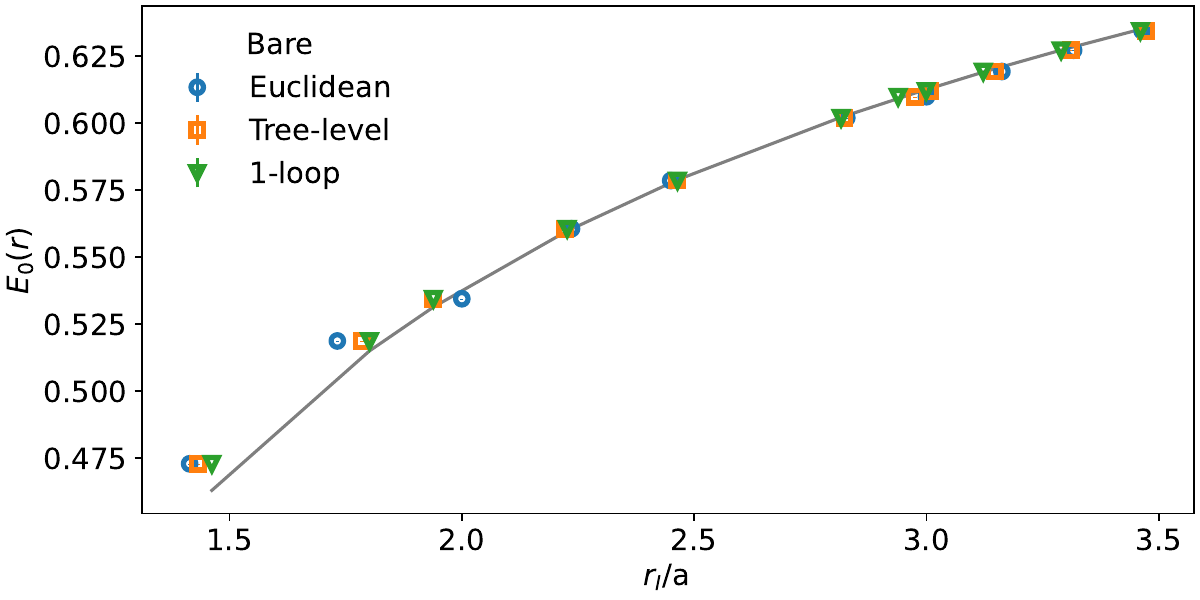}
    \includegraphics[width=0.49\textwidth]{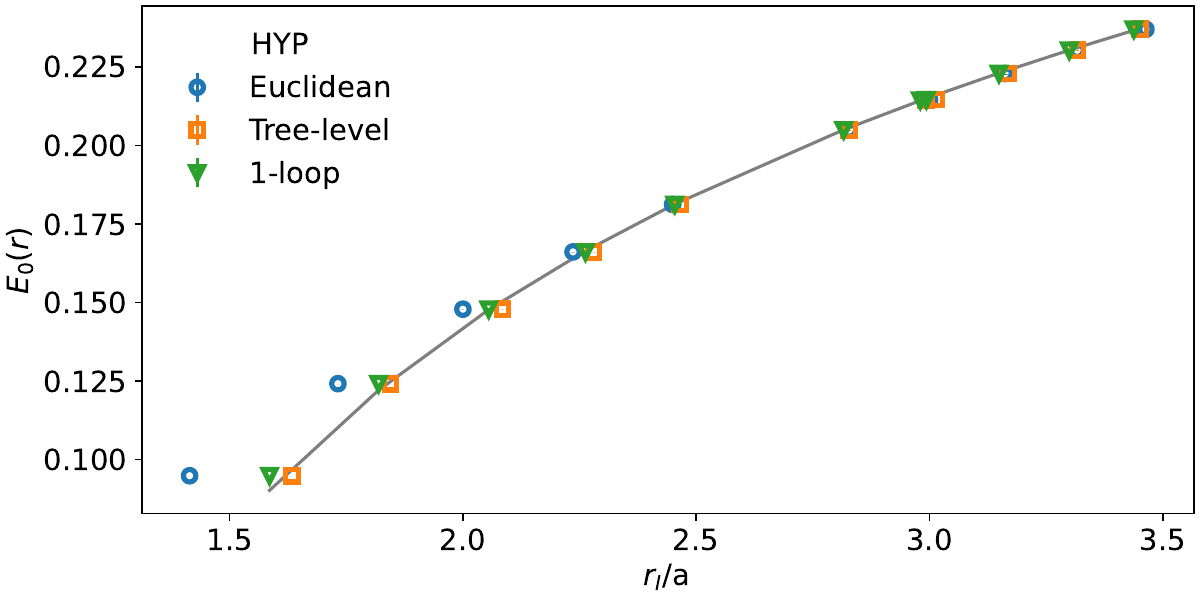}
    \vspace{-2.5cm}
    \begin{center}
    \hspace{3cm}\textcolor{preliminary}{\rotatebox{15}{PRELIMINARY}}\hspace{6cm}\textcolor{preliminary}{\rotatebox{-10}{PRELIMINARY}}
    \end{center}
    \vspace{0.2cm}
    \caption{Different orders of improved distance $r_I$ applied to the finest lattice ensemble for the bare (Left) and HYP-smeared (right) $E_0(r)$. The black curve shows a coulombic trend line {with fixed coupling}.
             }
    \label{fig:lpt}
\end{figure}
After the static {energy} has been computed, we correct for the discretization artifacts. 
At short distances of the order of the lattice spacing $a$, the static energy exhibits significant non-smooth discretization errors. 
These cutoff effects are known at the leading order and can be removed with a so called tree-level improvement prescription, 
where an improved distance $r_I$ is defined such that the continuum and lattice perturbation theories match at leading order.
While tree-level improvement allows one to reach shorter distances more smoothly than plain Euclidean $r/a$, it is not enough for the few first separations.
To extend the improvement to even shorter distances, we have performed a rigorous lattice perturbation theory calculation at next-to-leading order (one loop)~\cite{lpt_paper}. 
The lattice perturbation theory calculation is performed numerically with the \texttt{HiPPy} and \texttt{HPsrc} packages~\cite{Hart:2009nr}. 
The effect of the different levels of improvement is demonstrated in figure~\ref{fig:lpt} where the improvement is shown for both bare {or} HYP-smeared~\cite{Hasenfratz:2001hp} static quarks. 
In the figure, one can see that in the bare data for the two distinct path topologies at $r=3=\sqrt{3^2}=\sqrt{3^2+0^2+0^2}$ or~$=\sqrt{2^2+2^2+1^2}$, 
the one-loop correction is clearly making the overall shape smoother. 
The biggest correction from the one-loop improvement compared to the tree-level improvement happens at these on-axis path topologies where the Wilson lines are separated in single cardinal direction. 
The one-loop calculation in Ref.~\cite{lpt_paper} is still being finalized, and for these proceedings we stick with the tree-level improvement.
To compensate for the missing {1}-loop effects, we add $0.1$\% error as an extra systematics to all points $r^2\le 8a$, 
and to further reduce the unwanted discretization effects, we exclude the on-axis points from any of the fits.

\section{Extraction of \texorpdfstring{\boldmath$\Lambda_{\MSb}$}{Lambda MSbar}}
To extract $\Lambda_{\MSb}$, we perform a two parameter fit to the perturbative descriptions arising from the two approaches we use to regulate the renormalons.
The first fitted parameter is a shift constant $\Lambda$ that matches the perturbative curve to the lattice data at some distance $r^\ast$.
Secondly, we fit $\Lambda_{\MSb}$ that enters the equation~\eqref{eq:full_statenergy} via $\as(\mu=1/r,\Lambda_{\MSb})$. The running of the coupling is 
handled by the \texttt{RunDec}~\cite{Herren:2017osy} library with matching order of power counting to the static potential expansion. 

The perturbative formulas describe the non-perturbative lattice data only up to some distances.  
From previous TUMQCD extractions of $\Lambda_{\MSb}$~\cite{Bazavov:2012ka,Bazavov:2014soa,Bazavov:2019qoo}, we know that the perturbation theory works well up to $\sim 0.13$~fm. 
However, since the charm quark decouples around $0.15$~fm, the behavior at $0.13$~fm can already be affected by finite mass effects. 
Therefore, we perform two types of fits. Firstly, we restrict to very short distances $r_\mathrm{max} \simeq 0.1$~fm  and perform a fit with 4 massless quarks to extract $\Lambda_{\MSb}^{(4)}$. 
The caveat of this approach is that we are now more susceptible to the discretization errors. 
Secondly, we include the charm quark effects to the fit function, as described above, and fit up to separations $r_\mathrm{max} = 0.13$~fm with three massless quarks and one heavy quark to extract $\Lambda_{\MSb}^{(3)}$. 
The caveat of this approach is that the finite mass effects are known only up to two loops which restricts the entire analysis to two loop accuracy.
The third option would be to fit at large distances with 3 massless quarks, but for this to work we would need to fit in range $\gsim 0.18$~fm which is clearly out of the perturbative range and hence not considered.
The Lambda parameters with different number of fermions $\Lambda_{\MSb}^{(\Nf)}$ can be related to each other by perturbative decoupling relations. 
For consistency, and for easy comparison with the previous TUMQCD 2+1-flavor extractions, we present all the results as $\Lambda_{\MSb}^{(3)}$ in these proceedings

\begin{figure}
    \centering
    \begin{minipage}[t]{0.49\textwidth}\vspace{0pt}%
    \includegraphics[width=0.99\textwidth]{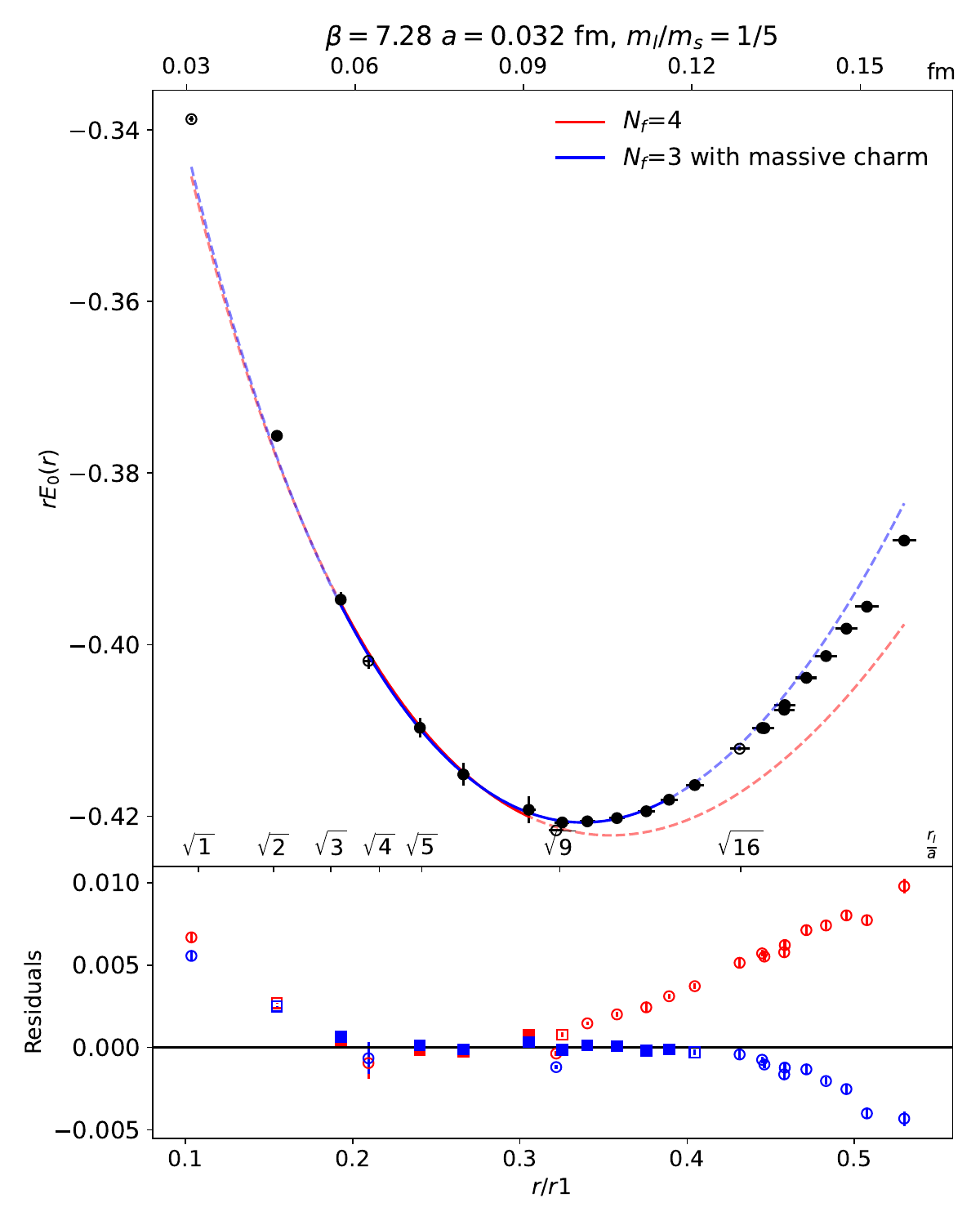}
    \vspace{-7cm}
    \begin{center}
    \hspace{-0cm}\textcolor{preliminary}{\rotatebox{-30}{PRELIMINARY}}
    \end{center}
    \vspace{3.75cm}
    \end{minipage}
    \hfill
    \begin{minipage}[t]{0.49\textwidth}\vspace{4ex}%
    \vspace{-0.42em}
    \includegraphics[width=1.01\textwidth]{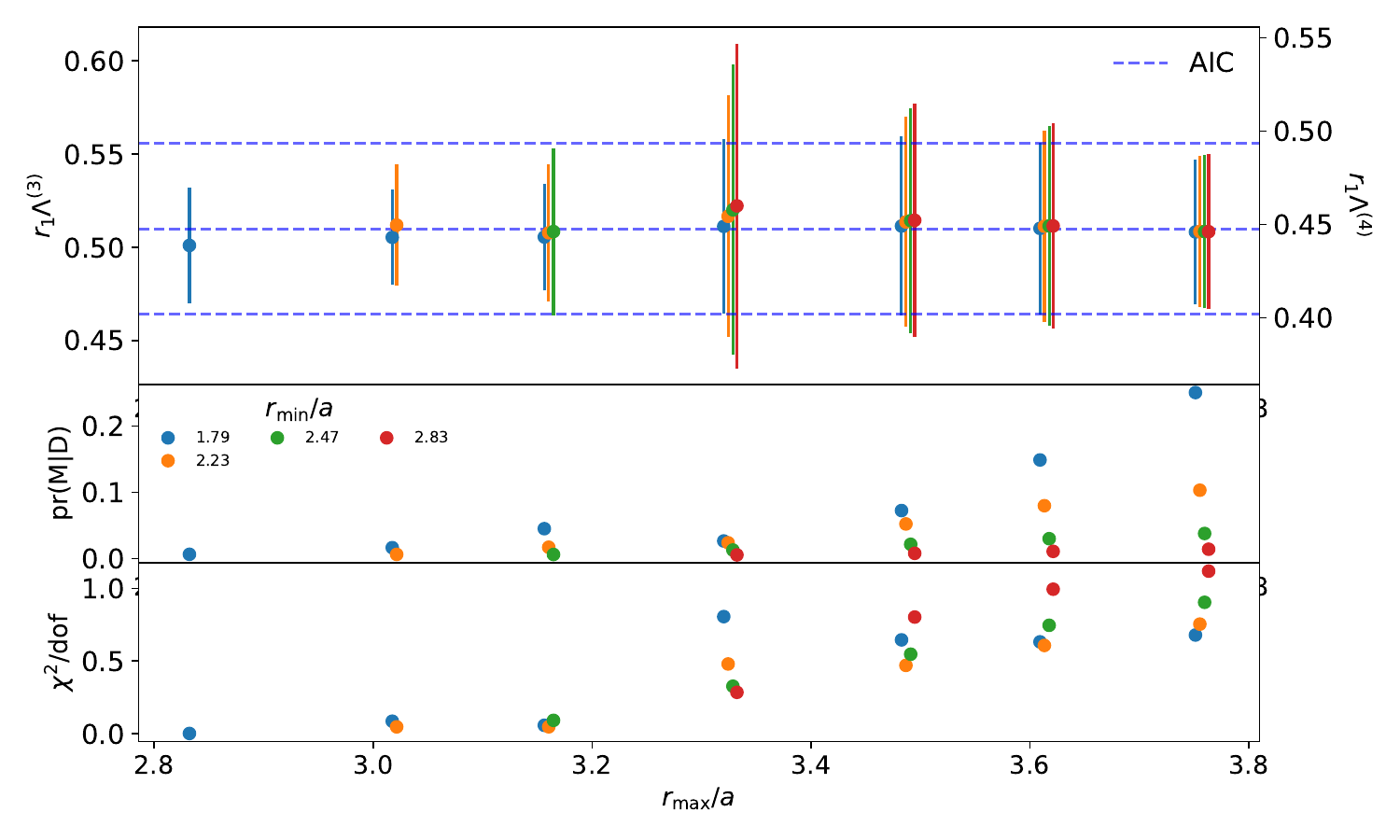}
    \caption{Left: Comparison of the lattice data to the perturbative curves with the fitted $\Lambda_{\MSb}$. 
             Right: the fit results for individual choices of $r_\mathrm{min}$ and $r_\mathrm{max}$ for the 3 flavor + charm fit (blue curve on the left) 
             together with the model weights and $\chi^2/\mathrm{d.o.f}$
             }
    \label{fig:lambdafit}
    \end{minipage}
\end{figure}
With these limits, we perform correlated fits to data divided into 100 jackknife blocks and fit to all possible ranges $a< r_\mathrm{min} \le r \le r_\mathrm{max}$ that have at least three data points. 
The fits from different ranges are then combined with a model average using the Akaike information criterion (AIC) to {weigh} the different fits~\cite{Jay:2020jkz}. 
Since only small fractions of the fits have non-zero AIC weight and since these ranges tend to be same between different jackknife blocks, we make sure to use {the} same set of selected fit ranges for all jackknife blocks.

\begin{figure}
    \centering
    \includegraphics[width=0.49\textwidth]{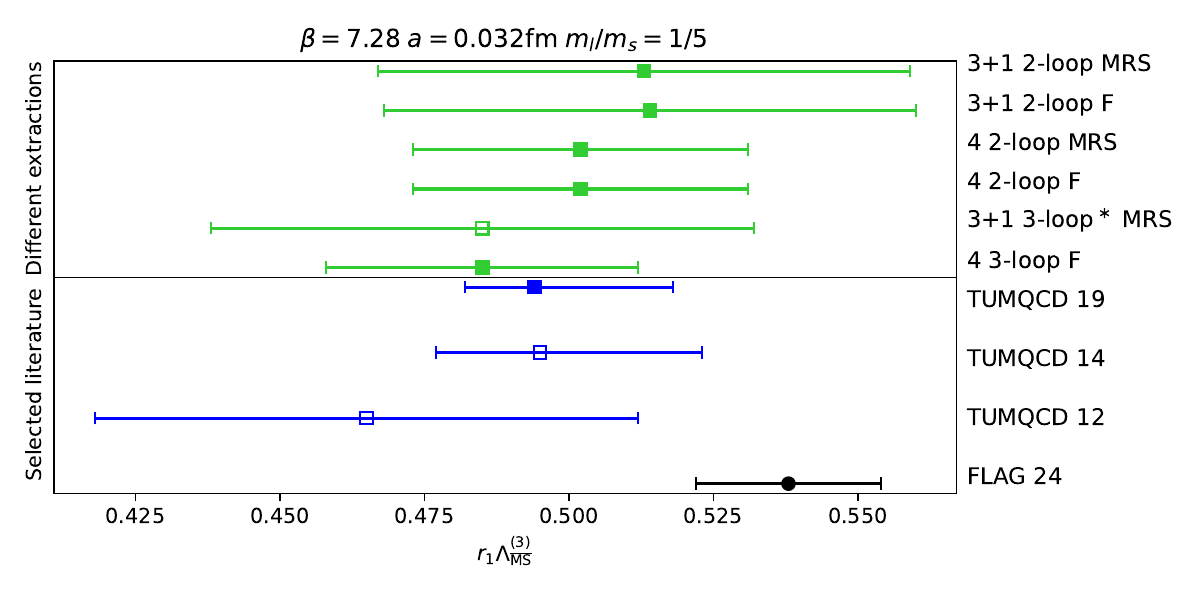}
    \includegraphics[width=0.49\textwidth]{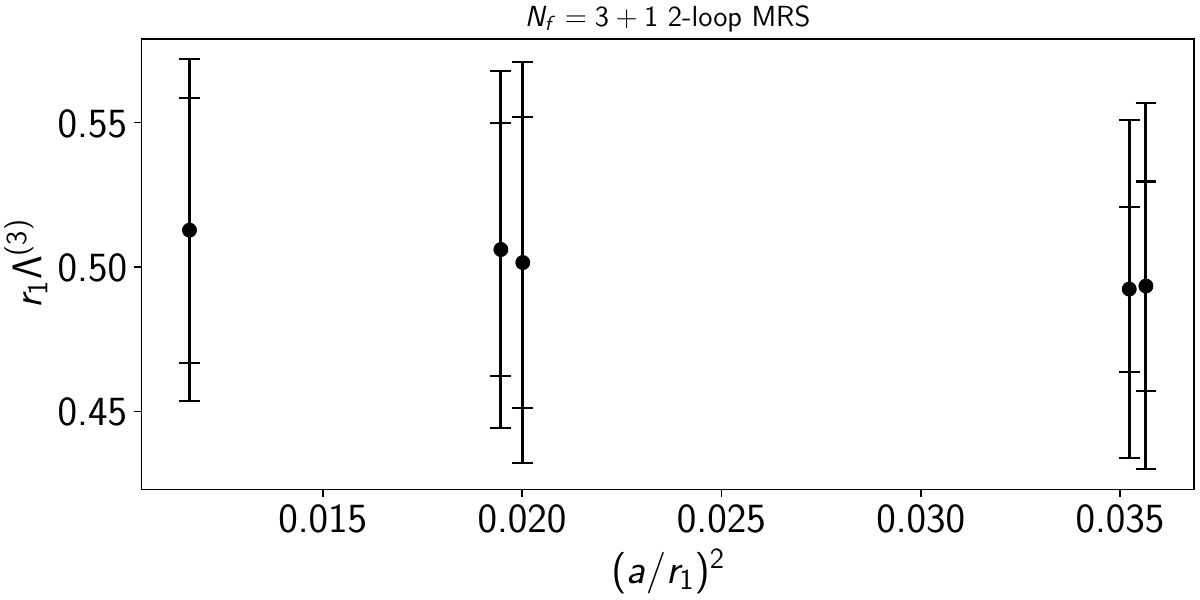}
    \vspace{-2.5cm}
    \begin{center}
    \hspace{+10cm}\textcolor{preliminary}{\rotatebox{-15}{PRELIMINARY}}
    \end{center}
    \vspace{0.5cm}
    \caption{Left:  Resulting $r_1\Lambda_{\MSb}$ from the fits to the finest lattice ensemble at different orders of perturbative expansion in green. 
             The $\text{3-loop}^\ast$ point is incomplete at three~loops.
             In blue we show the previous TUMQCD extractions from 2+1 flavors and in black the current FLAG average. The FLAG point is calculated from their $r_0\Lambda$ by multiplying it by the FLAG reported ratio $r_1/r_0$. 
             Right: One of the fits from the left hand side figure performed to coarser ensembles. The points close to each other have the same lattice spacing (shifted for visibility) but different light quark mass. 
             The inner error bar indicates the statistical error, while the outer error bar shows the systematic error from the variation of the soft scale $\mu=1/r$ by a factor of $\sqrt{2}$.
             }
    \label{fig:results}
\end{figure}

This fit procedure is demonstrated for the finest ensemble in figure~\ref{fig:lambdafit}.
The black points show the lattice data, with the filled points being included in the fit and the empty on-axis point being excluded.
The two curves present the fits either with or without the massive charm effects and the solid region represents the range of the fit with best AIC weight. 
Similarly, on the bottom half of the figure, the filled points correspond to the solid sections of respective lines from above. 
We observe that the fits describe the data well within the range.
Apart from doing the fits with $\Nf=4$ and $\Nf=3+1$ flavors, we also repeat the analysis with different orders of {the} perturbative expansion of the static energy and using both renormalon {removal} methods.
A collection of these fit results for the finest ensemble is shown on the left side of figure~\ref{fig:results}. 
Since the finite mass corrections are known only at two-loop level, the extraction with the three-loop static energy with massive charm is incomplete at three loops. 
Nevertheless, we observe stability between all fit options and general agreement 
with previous TUMQCD 2+1-flavor extractions~\cite{Bazavov:2012ka,Bazavov:2014soa,Bazavov:2019qoo} and the most recent FLAG average~\cite{Aoki:2024oxs}.

The fit is then repeated for different scaling factors $c$ of the soft scale $\mu=c/r$. Ideally, one would vary the scale by a factor of two. 
However, as can be seen from the left of figure~\ref{fig:pert}, the scale variation $c=1/2$ will heavily constraint the available $r$-values for the fit.
Therefore, at the time of the conference, we could only afford to vary the scale within a more relaxed range $c\in \{1/\sqrt{2},\sqrt{2}\}$, due to the limited range in $r$ and from the missing one-loop improvement.
Moreover, the fit is then repeated over all ensembles. 
Again, due to available fit ranges with the chosen range of scale variation $c$, we could only extract $\Lambda_{\MSb}$ at three finest lattice spacings with {a few} different light quark masses.
Using the $\Nf=3+1$ two-loop MRS fit as a reference, we show the lattice spacing dependence of the extracted $\Lambda_{\MSb}^{(3)}$ on the right side of figure~\ref{fig:results}, 
where the points that are grouped together have the same lattice spacing $a$ but different light quark mass ratio $m_l/m_s$. 
We observe mild dependency on the lattice spacing and light quark masses, however, the proper continuum extrapolation is left for future publication. 
As the results in these proceedings are not final, we only present them in visualized form and refrain from giving quotable numbers until the final publication.
Once the $\Lambda_{\MSb}$-parameter has been determined, we can use \texttt{RunDec} to obtain the strong coupling at the Z-pole mass $\as(M_\mathrm{Z})$.

\subsection{Physical scale}\label{sec:scale}
In these proceedings we decided to present results in $r_1$-units. 
In order to convert to physical units, we can use our scale determination~\cite{Brambilla:2022het} $r_1=0.3037(25)$~fm, which is based on the $a_{f_{p4s}}$ quantity discussed in Ref.~\cite{MILC:2012znn}.
The previous TUMQCD extraction with 2+1-flavors~\cite{Bazavov:2019qoo} used a different value for the scale: $r_1 = 0.3106(17)$~fm based on $f_\pi$ scale.
This means that if the comparison on the left side of figure~\ref{fig:results} was presented in physical units, there would be slightly more pronounced difference between the $\Lambda_{\MSb}^{(3)}$ extracted from the {(2+1+1)- or (2+1)-flavor} datasets.
In an upcoming study~\cite{JohannesPeter:2025}, a new determination of {(2+1)}-flavor $r_1$ will be presented based on the Kaon decay constant $f_\mathrm{K}$.
This new constant moves the $r_1$ value closer to the {(2+1+1)}-flavor $r_1$ by a few standard deviations. 
In general, one does not expect there to be significant finite mass effects in the $r_1$ scale and hence the three and four flavor physical scales should be close to each other.

\acknowledgments{
The lattice QCD calculations have been performed using the publicly available \href{https://web.physics.utah.edu/~detar/milc/milcv7.html}{MILC code}. 
The simulations were carried out on the computing facilities of the Computational Center for Particle and Astrophysics (C2PAP) in the project 
\emph{Calculation of finite $T$ QCD correlators} (pr83pu) and of the SuperMUC cluster at the Leibniz-Rechenzentrum (LRZ) in the project
\emph{The role of the charm-quark for the QCD coupling constant} (pn56bo),
both located in Munich (Germany),
The authors acknowledge the Gauss Centre for Supercomputing e.V.
(\href{www.gauss-centre.eu}{www.gauss-centre.eu})
for funding this project by providing computing time on the GCS Supercomputer SuperMUC-NG
at Leibniz Supercomputing Centre (\href{www.lrz.de}{www.lrz.de}).
This research was funded by the Deutsche Forschungsgemeinschaft (DFG, German Research Foundation) cluster of excellence "ORIGINS" (\href{https://www.origins-cluster.de}{www.origins-cluster.de}) under Germany's Excellence Strategy EXC-2094-390783311.
J.H.W.’s research has been funded by the Deutsche Forschungsgemeinschaft (DFG, German Research Foundation)---Projektnummer 417533893/GRK2575 ``Rethinking Quantum Field Theory''.  The authors acknowledge the support by the State of Hesse within the Research Cluster ELEMENTS (Project ID 500/10.006)
N. B. acknowledges the European Research Council advanced grant ERC-2023-ADG-Project EFT-XYZ.
P.P. was supported by The U.S. Department of Energy through Contract No.~DE-SC0012704.
A.~B.'s work was supported by the US National Science Foundation under grant No. PHY23-09946.
This document was prepared by TUMQCD using the resources of the Fermi National Accelerator Laboratory (Fermilab), a U.S. Department of Energy, Office of Science, Office of High Energy Physics HEP User Facility. Fermilab is managed by Fermi Forward Discovery Group, LLC, acting under Contract No.\ 89243024CSC000002.}

\bibliographystyle{jhep_modified}
\bibliography{alphaS.bib}

\providecommand{\href}[2]{#2}\begingroup\raggedright\begin{thebibliography}{10}

\bibitem{Bali:2000gf}
G.~S. Bali, \emph{{QCD} forces and heavy quark bound states},
  \href{https://doi.org/10.1016/S0370-1573(00)00079-X}{\emph{Phys. Rept.}
  {\bfseries 343} (2001) 1}
  [\href{https://arxiv.org/abs/hep-ph/0001312}{{\ttfamily hep-ph/0001312}}].
\bibitem{Brambilla:1999qa}
N.~Brambilla, A.~Pineda, J.~Soto and A.~Vairo, \emph{The infrared behavior of
  the static potential in perturbative {QCD}},
  \href{https://doi.org/10.1103/PhysRevD.60.091502}{\emph{Phys. Rev. D}
  {\bfseries 60} (1999) 091502}
  [\href{https://arxiv.org/abs/hep-ph/9903355}{{\ttfamily hep-ph/9903355}}].
\bibitem{Pineda:2000gza}
A.~Pineda and J.~Soto, \emph{The renormalization group improvement of the {QCD}
  static potentials},
  \href{https://doi.org/10.1016/S0370-2693(00)01261-2}{\emph{Phys. Lett. B}
  {\bfseries 495} (2000) 323}
  [\href{https://arxiv.org/abs/hep-ph/0007197}{{\ttfamily hep-ph/0007197}}].
\bibitem{Brambilla:2006wp}
N.~Brambilla, X.~Garcia~i Tormo, J.~Soto and A.~Vairo, \emph{The logarithmic
  contribution to the {QCD} static energy at {N$^{4}$LO}},
  \href{https://doi.org/10.1016/j.physletb.2007.02.015}{\emph{Phys. Lett. B}
  {\bfseries 647} (2007) 185}
  [\href{https://arxiv.org/abs/hep-ph/0610143}{{\ttfamily hep-ph/0610143}}].
\bibitem{Brambilla:2009bi}
N.~Brambilla, A.~Vairo, X.~Garcia~i Tormo and J.~Soto, \emph{The {QCD} static
  energy at {N$^{3}$LL}},
  \href{https://doi.org/10.1103/PhysRevD.80.034016}{\emph{Phys. Rev. D}
  {\bfseries 80} (2009) 034016}
  [\href{https://arxiv.org/abs/0906.1390}{{\ttfamily 0906.1390}}].
\bibitem{Anzai:2009tm}
C.~Anzai, Y.~Kiyo and Y.~Sumino, \emph{Static {QCD} potential at three-loop
  order}, \href{https://doi.org/10.1103/PhysRevLett.104.112003}{\emph{Phys.
  Rev. Lett.} {\bfseries 104} (2010) 112003}
  [\href{https://arxiv.org/abs/0911.4335}{{\ttfamily 0911.4335}}].
\bibitem{Smirnov:2009fh}
A.~V. Smirnov, V.~A. Smirnov and M.~Steinhauser, \emph{Three-loop static
  potential}, \href{https://doi.org/10.1103/PhysRevLett.104.112002}{\emph{Phys.
  Rev. Lett.} {\bfseries 104} (2010) 112002}
  [\href{https://arxiv.org/abs/0911.4742}{{\ttfamily 0911.4742}}].
\bibitem{Brambilla:2010pp}
N.~Brambilla, X.~Garcia~i Tormo, J.~Soto and A.~Vairo, \emph{{Precision
  determination of $r_0\Lambda_{\overline{\rm MS}}$ from the QCD static
  energy}}, \href{https://doi.org/10.1103/PhysRevLett.105.212001}{\emph{Phys.
  Rev. Lett.} {\bfseries 105} (2010) 212001}
  [\href{https://arxiv.org/abs/1006.2066}{{\ttfamily 1006.2066}}].
\bibitem{Husung:2017qjz}
N.~Husung, M.~Koren, P.~Krah and R.~Sommer, \emph{{SU(3) Yang Mills theory at
  small distances and fine lattices}},
  \href{https://doi.org/10.1051/epjconf/201817514024}{\emph{EPJ Web Conf.}
  {\bfseries 175} (2018) 14024}
  [\href{https://arxiv.org/abs/1711.01860}{{\ttfamily 1711.01860}}].
\bibitem{Brambilla:2023fsi}
N.~Brambilla, V.~Leino, J.~Mayer-Steudte and A.~Vairo, \emph{{Static force from
  generalized Wilson loops on the lattice using the gradient flow}},
  \href{https://doi.org/10.1103/PhysRevD.109.114517}{\emph{Phys. Rev. D}
  {\bfseries 109} (2024) 114517}
  [\href{https://arxiv.org/abs/2312.17231}{{\ttfamily 2312.17231}}].
\bibitem{Jansen:2011vv}
{\scshape ETM} collaboration, K.~Jansen, F.~Karbstein, A.~Nagy and M.~Wagner,
  \emph{{$\Lambda_{\overline{\rm MS}}$ from the static potential for QCD with
  $n_f=2$ dynamical quark flavors}},
  \href{https://doi.org/10.1007/JHEP01(2012)025}{\emph{JHEP} {\bfseries 01}
  (2012) 025} [\href{https://arxiv.org/abs/1110.6859}{{\ttfamily 1110.6859}}].
\bibitem{Karbstein:2014bsa}
F.~Karbstein, A.~Peters and M.~Wagner,
  \emph{{${\Lambda}_{\overline{\mathrm{MS}}}^{({n}_f=2)}$ from a momentum space
  analysis of the quark-antiquark static potential}},
  \href{https://doi.org/10.1007/JHEP09(2014)114}{\emph{JHEP} {\bfseries 09}
  (2014) 114} [\href{https://arxiv.org/abs/1407.7503}{{\ttfamily 1407.7503}}].
\bibitem{Karbstein:2018mzo}
F.~Karbstein, M.~Wagner and M.~Weber, \emph{{Determination of
  $\Lambda_{\overline{\textrm{MS}}}^{(n_f=2)}$ and analytic parametrization of
  the static quark-antiquark potential}},
  \href{https://doi.org/10.1103/PhysRevD.98.114506}{\emph{Phys. Rev. D}
  {\bfseries 98} (2018) 114506}
  [\href{https://arxiv.org/abs/1804.10909}{{\ttfamily 1804.10909}}].
\bibitem{Bazavov:2012ka}
A.~Bazavov, N.~Brambilla, X.~Garcia~i Tormo, P.~Petreczky, J.~Soto and
  A.~Vairo, \emph{{Determination of $\alpha_s$ from the QCD static energy}},
  \href{https://doi.org/10.1103/PhysRevD.86.114031}{\emph{Phys. Rev. D}
  {\bfseries 86} (2012) 114031}
  [\href{https://arxiv.org/abs/1205.6155}{{\ttfamily 1205.6155}}].
\bibitem{Bazavov:2014soa}
A.~Bazavov, N.~Brambilla, X.~Garcia~i Tormo, P.~Petreczky, J.~Soto and
  A.~Vairo, \emph{Determination of {$\alpha_{\text{s}}$} from the {QCD} static
  energy: An update},
  \href{https://doi.org/10.1103/PhysRevD.90.074038}{\emph{Phys. Rev. D}
  {\bfseries 90} (2014) 074038}
  [\href{https://arxiv.org/abs/1407.8437}{{\ttfamily 1407.8437}}].
\bibitem{Takaura:2018vcy}
H.~Takaura, T.~Kaneko, Y.~Kiyo and Y.~Sumino, \emph{Determination of
  {$\alpha_{\text{s}}$} from static {QCD} potential: {OPE} with renormalon
  subtraction and lattice {QCD}},
  \href{https://doi.org/10.1007/JHEP04(2019)155}{\emph{JHEP} {\bfseries 04}
  (2019) 155} [\href{https://arxiv.org/abs/1808.01643}{{\ttfamily
  1808.01643}}].
\bibitem{Bazavov:2019qoo}
{\scshape TUMQCD} collaboration, A.~Bazavov, N.~Brambilla, X.~Garcia~i Tormo,
  P.~Petreczky, J.~Soto, A.~Vairo et~al., \emph{Determination of the {QCD}
  coupling from the static energy and the free energy},
  \href{https://doi.org/10.1103/PhysRevD.100.114511}{\emph{Phys. Rev. D}
  {\bfseries 100} (2019) 114511}
  [\href{https://arxiv.org/abs/1907.11747}{{\ttfamily 1907.11747}}].
\bibitem{Ayala:2020odx}
C.~Ayala, X.~Lobregat and A.~Pineda, \emph{Determination of {$\alpha(M_{Z})$}
  from an hyperasymptotic approximation to the energy of a static
  quark-antiquark pair},
  \href{https://doi.org/10.1007/JHEP09(2020)016}{\emph{JHEP} {\bfseries 09}
  (2020) 016} [\href{https://arxiv.org/abs/2005.12301}{{\ttfamily
  2005.12301}}].
\bibitem{Blossier:2013ioa}
{\scshape ETM} collaboration, B.~Blossier, P.~Boucaud, M.~Brinet, F.~De~Soto,
  V.~Morenas, O.~Pene et~al., \emph{{High statistics determination of the
  strong coupling constant in Taylor scheme and its OPE Wilson coefficient from
  lattice QCD with a dynamical charm}},
  \href{https://doi.org/10.1103/PhysRevD.89.014507}{\emph{Phys. Rev. D}
  {\bfseries 89} (2014) 014507}
  [\href{https://arxiv.org/abs/1310.3763}{{\ttfamily 1310.3763}}].
\bibitem{Chakraborty:2014aca}
{\scshape HPQCD} collaboration, B.~Chakraborty, C.~T.~H. Davies, B.~Galloway,
  P.~Knecht, J.~Koponen, G.~C. Donald et~al., \emph{High-precision quark masses
  and {QCD} coupling from {$N_{\text{f}} = 4$} lattice {QCD}},
  \href{https://doi.org/10.1103/PhysRevD.91.054508}{\emph{Phys. Rev. D}
  {\bfseries 91} (2015) 054508}
  [\href{https://arxiv.org/abs/1408.4169}{{\ttfamily 1408.4169}}].
\bibitem{Aoki:2024oxs}
{\scshape Flavour Lattice Averaging Group (FLAG)} collaboration, Y.~Aoki
  et~al., \emph{{FLAG Review 2024}},
  \href{https://arxiv.org/abs/2411.04268}{{\ttfamily 2411.04268}}.
\bibitem{dEnterria:2022hzv}
D.~d'Enterria et~al., \emph{{The strong coupling constant: State of the art and
  the decade ahead}}, \href{https://doi.org/10.1088/1361-6471/ad1a78}{\emph{J.
  Phys. G} {\bfseries 51} (2024) 090501}
  [\href{https://arxiv.org/abs/2203.08271}{{\ttfamily 2203.08271}}].
\bibitem{Tormo:2013tha}
X.~Garcia~i Tormo, \emph{Review on the determination of {$\alpha_{\text{s}}$}
  from the {QCD} static energy},
  \href{https://doi.org/10.1142/S0217732313300280}{\emph{Mod. Phys. Lett. A}
  {\bfseries 28} (2013) 1330028}
  [\href{https://arxiv.org/abs/1307.2238}{{\ttfamily 1307.2238}}].
\bibitem{Recksiegel:2001xq}
S.~Recksiegel and Y.~Sumino, \emph{Perturbative {QCD} potential, renormalon
  cancellation and phenomenological potentials},
  \href{https://doi.org/10.1103/PhysRevD.65.054018}{\emph{Phys. Rev. D}
  {\bfseries 65} (2002) 054018}
  [\href{https://arxiv.org/abs/hep-ph/0109122}{{\ttfamily hep-ph/0109122}}].
\bibitem{Brambilla:2021wqs}
N.~Brambilla, V.~Leino, O.~Philipsen, C.~Reisinger, A.~Vairo and M.~Wagner,
  \emph{{Lattice gauge theory computation of the static force}},
  \href{https://doi.org/10.1103/PhysRevD.105.054514}{\emph{Phys. Rev. D}
  {\bfseries 105} (2022) 054514}
  [\href{https://arxiv.org/abs/2106.01794}{{\ttfamily 2106.01794}}].
\bibitem{Brambilla:2017hcq}
{\scshape TUMQCD} collaboration, N.~Brambilla, J.~Komijani, A.~S. Kronfeld and
  A.~Vairo, \emph{Relations between heavy-light meson and quark masses},
  \href{https://doi.org/10.1103/PhysRevD.97.034503}{\emph{Phys. Rev. D}
  {\bfseries 97} (2018) 034503}
  [\href{https://arxiv.org/abs/1712.04983}{{\ttfamily 1712.04983}}].
\bibitem{Komijani:2017vep}
J.~Komijani, \emph{{A discussion on leading renormalon in the pole mass}},
  \href{https://doi.org/10.1007/JHEP08(2017)062}{\emph{JHEP} {\bfseries 08}
  (2017) 062} [\href{https://arxiv.org/abs/1701.00347}{{\ttfamily
  1701.00347}}].
\bibitem{Kronfeld:2023jab}
A.~S. Kronfeld, \emph{{Factorial growth at low orders in perturbative QCD:
  control over truncation uncertainties}},
  \href{https://doi.org/10.1007/JHEP12(2023)108}{\emph{JHEP} {\bfseries 12}
  (2023) 108} [\href{https://arxiv.org/abs/2310.15137}{{\ttfamily
  2310.15137}}].
\bibitem{MILC:2010pul}
{\scshape MILC} collaboration, A.~Bazavov et~al., \emph{Scaling studies of
  {QCD} with the dynamical {HISQ} action},
  \href{https://doi.org/10.1103/PhysRevD.82.074501}{\emph{Phys. Rev. D}
  {\bfseries 82} (2010) 074501}
  [\href{https://arxiv.org/abs/1004.0342}{{\ttfamily 1004.0342}}].
\bibitem{MILC:2012znn}
{\scshape MILC} collaboration, A.~Bazavov et~al., \emph{Lattice {QCD} ensembles
  with four flavors of highly improved staggered quarks},
  \href{https://doi.org/10.1103/PhysRevD.87.054505}{\emph{Phys. Rev. D}
  {\bfseries 87} (2013) 054505}
  [\href{https://arxiv.org/abs/1212.4768}{{\ttfamily 1212.4768}}].
\bibitem{Bazavov:2017lyh}
{\scshape Fermilab Lattice, MILC} collaboration, A.~Bazavov et~al.,
  \emph{{$B$}- and {$D$}-meson leptonic decay constants from four-flavor
  lattice {QCD}}, \href{https://doi.org/10.1103/PhysRevD.98.074512}{\emph{Phys.
  Rev. D} {\bfseries 98} (2018) 074512}
  [\href{https://arxiv.org/abs/1712.09262}{{\ttfamily 1712.09262}}].
\bibitem{Follana:2006rc}
{\scshape HPQCD} collaboration, E.~Follana, Q.~Mason, C.~Davies, K.~Hornbostel,
  G.~P. Lepage, J.~Shigemitsu et~al., \emph{Highly improved staggered quarks on
  the lattice, with applications to charm physics},
  \href{https://doi.org/10.1103/PhysRevD.75.054502}{\emph{Phys. Rev. D}
  {\bfseries 75} (2007) 054502}
  [\href{https://arxiv.org/abs/hep-lat/0610092}{{\ttfamily hep-lat/0610092}}].
\bibitem{Brambilla:2022het}
{\scshape TUMQCD} collaboration, N.~Brambilla, R.~L. Delgado, A.~S. Kronfeld,
  V.~Leino, P.~Petreczky, S.~Steinbei{\ss}er et~al., \emph{{Static energy in
  ($2+1+1$)-flavor lattice QCD: Scale setting and charm effects}},
  \href{https://doi.org/10.1103/PhysRevD.107.074503}{\emph{Phys. Rev. D}
  {\bfseries 107} (2023) 074503}
  [\href{https://arxiv.org/abs/2206.03156}{{\ttfamily 2206.03156}}].
\bibitem{Bernard:2000gd}
C.~W. Bernard, T.~Burch, K.~Orginos, D.~Toussaint, T.~A. DeGrand, C.~E. DeTar
  et~al., \emph{Static quark potential in three-flavor {QCD}},
  \href{https://doi.org/10.1103/PhysRevD.62.034503}{\emph{Phys. Rev. D}
  {\bfseries 62} (2000) 034503}
  [\href{https://arxiv.org/abs/hep-lat/0002028}{{\ttfamily hep-lat/0002028}}].
\bibitem{lpt_paper}
G.~M. von Hippel, V.~Leino and S.~Steinbeißer, \emph{{One loop improvement of
  the static potential with HISQ quarks}}, {\emph{In preparation: TUM-EFT
  171/22} (2025) }.
\bibitem{Hart:2009nr}
A.~Hart, G.~M. von Hippel, R.~R. Horgan and E.~H. Muller, \emph{Automated
  generation of lattice {QCD} feynman rules},
  \href{https://doi.org/10.1016/j.cpc.2009.04.021}{\emph{Comput. Phys. Commun.}
  {\bfseries 180} (2009) 2698}
  [\href{https://arxiv.org/abs/0904.0375}{{\ttfamily 0904.0375}}].
\bibitem{Hasenfratz:2001hp}
A.~Hasenfratz and F.~Knechtli, \emph{Flavor symmetry and the static potential
  with hypercubic blocking},
  \href{https://doi.org/10.1103/PhysRevD.64.034504}{\emph{Phys. Rev. D}
  {\bfseries 64} (2001) 034504}
  [\href{https://arxiv.org/abs/hep-lat/0103029}{{\ttfamily hep-lat/0103029}}].
\bibitem{Herren:2017osy}
F.~Herren and M.~Steinhauser, \emph{{Version~3 of \texttt{RunDec} and
  \texttt{CRunDec}}},
  \href{https://doi.org/10.1016/j.cpc.2017.11.014}{\emph{Comput. Phys. Commun.}
  {\bfseries 224} (2018) 333}
  [\href{https://arxiv.org/abs/1703.03751}{{\ttfamily 1703.03751}}].
\bibitem{Jay:2020jkz}
W.~I. Jay and E.~T. Neil, \emph{{Bayesian} model averaging for analysis of
  lattice field theory results},
  \href{https://doi.org/10.1103/PhysRevD.103.114502}{\emph{Phys. Rev. D}
  {\bfseries 103} (2021) 114502}
  [\href{https://arxiv.org/abs/2008.01069}{{\ttfamily 2008.01069}}].
\bibitem{JohannesPeter:2025}
P.~Petreczky and J.~H. Weber, \emph{\textit{in preparation}}, {\emph{private
  communication} (2025) }.
\end{thebibliography}\endgroup

\end{document}